\documentclass{article}
 
\usepackage{microtype}
\usepackage{graphicx}
\usepackage{subfigure}
\usepackage{booktabs} 

\usepackage{hyperref}

\usepackage[accepted]{icml2025}

\usepackage{amsmath}
\usepackage{amssymb}
\usepackage{mathtools}
\usepackage{amsthm}

\usepackage[capitalize,noabbrev]{cleveref}

\theoremstyle{plain}

\theoremstyle{definition}

\theoremstyle{remark}

\usepackage[textsize=tiny]{todonotes}

\icmltitlerunning{Geometric SSM: LTI State Space Models for Selective Tasks}

\usepackage{multirow}
\usepackage{stfloats}
\usepackage{enumitem}
\usetikzlibrary {chains,scopes,positioning,decorations.pathreplacing,calc,fit} 
\tikzset{token/.style = {draw=none, ultra thin, minimum size=5mm},
trigger/.style = {draw=none,fill = black,ultra thin, minimum size=5mm},
blank/.style = {draw, ultra thin, minimum size=5mm},
questionMark/.style = {draw, ultra thin, minimum size=5mm},
braced/.style={
     decoration={brace, mirror},
     decorate
   },
   braceu/.style={
     decoration=brace,
     decorate
   },
   dot/.style={
   	circle,inner sep=0pt,minimum size=0pt,fill=white
   	},
    block/.style={draw, fill=white, rectangle, 
            minimum height=2em, minimum width=3em},
    blockSmall/.style={draw, fill=white, rectangle, 
            minimum height=1.25em, minimum width=1.45em},
    input/.style={inner sep=0pt},       
    output/.style={inner sep=0pt},      
    sum/.style = {draw, fill=white, circle, minimum size = 0.1cm, node distance=1.5cm, inner sep=0pt},
    cross/.style={path picture={\draw[black] (path picture bounding box.south east) -- (path picture bounding box.north west) (path picture bounding box.south west) -- (path picture bounding box.north east);
}},
    mult/.style = {draw, fill=white, circle,cross, minimum size = 0.125cm, node distance=1.5cm, inner sep=0pt},
    pinstyle/.style = {pin edge={to-,thin,black}},
    sigmoid/.style = {blockSmall, path picture ={
    \draw[shift={(path picture bounding box.center)},
        domain=-0.3:0.3, samples=50,
    ] plot ({\x},{0.3*(sigma(30*\x)-0.5)});
}},
    declare function={
        sigma(\x) = 1/(1+exp(-\x));
    }
}
\definecolor{myBlue}{rgb}{0,0.4470,0.7410}
\definecolor{myRed}{rgb}{0.8500,0.3250, 0.0980}
\definecolor{myYellow}{rgb}{0.9290,0.6940,0.1250}
\definecolor{myPurple}{rgb}{0.4940, 0.1840, 0.5560}
\definecolor{myGreen}{rgb}{0.4660,0.6740,0.1880}

\def\nf{n_f}
\def\nM{n_M}
\def\nRes{n_r}
\def\RR{\mathbb{R}}
\def\Rm{{\mathbb{R}^m}}
\def\ROnem{{\mathbb{R}^{1\times m}}}

\def\Rnfnf{{\mathbb{R}^{\nf\times \nf}}}
\def\Rnfm{{\mathbb{R}^{\nf\times m}}}
\def\Rmnf{{\mathbb{R}^{ m \times \nf}}}

\def\RnResnRes{{\mathbb{R}^{\nRes\times \nRes}}}
\def\RnResm{{\mathbb{R}^{\nRes\times m}}}
\def\ROnenRes{{\mathbb{R}^{ 1 \times \nRes}}}

\def\RmnM{{\mathbb{R}^{m\times \nM}}}
\def\RnMnM{{\mathbb{R}^{\nM\times \nM}}}
\def\RnMm{{\mathbb{R}^{\nM\times m}}}

\def\RnResnRes{{\mathbb{R}^{\nRes\times \nRes}}}
\def\RnResm{{\mathbb{R}^{\nRes\times m}}}

\def\Rnn{{\mathbb{R}^{n\times n}}}
\def\Rnm{{\mathbb{R}^{n\times m}}}
\def\Rmm{{\mathbb{R}^{m\times m}}}

\def\Ab{\overline{A}}
\def\Bb{\overline{B}}
\def\Cb{\overline{C}}

\def\Sf{\Sigma_{f}}
\def\rt{r{\left(t\right)}}

\def\ut{u{\left(t\right)}}
\def\uit{u_i{\left(t\right)}}
\def\yit{y_i{\left(t\right)}}

\def\yst{y_s{\left(t\right)}}
\def\st{s{\left(t\right)}}

\def\ft{f{\left(t\right)}}

\def\htime{h{\left(t\right)}}
\def\hitime{h^i{\left(t\right)}}

\def\hrtime{h_r{\left(t\right)}}
\def\hrtOne{h_r{\left(t+1\right)}}

\def\hftime{h_f{\left(t\right)}}
\def\hftOne{h_f{\left(t+1\right)}}

\def\hMtime{h_M{\left(t\right)}}
\def\hMtOne{h_M{\left(t+1\right)}}

\def\hitOne{h^i{\left(t+1\right)}}

\def\yt{y{\left(t\right)}}
\def\ytOne{y{\left(t+1\right)}}

\newcommand\oprocendsymbol{\hbox{$\square$}}
\newcommand\oprocend{\relax\ifmmode\else\unskip\hfill\fi\oprocendsymbol}

\begin{document}

\twocolumn[
\icmltitle{Geometric SSMs with LTI Dynamics for Selective Sequence Modeling}



\icmlsetsymbol{equal}{*}

\begin{icmlauthorlist}
\icmlauthor{Umberto Casti}{xxx}
\icmlauthor{Giacomo Baggio}{xxx}
\icmlauthor{Sandro Zampieri}{xxx}
\icmlauthor{Fabio Pasqualetti}{yyy}
\end{icmlauthorlist}

\icmlaffiliation{xxx}{Department of Information Engineering, University of Padova, Padova, Italy}
\icmlaffiliation{yyy}{Department of Mechanical Engineering, University of California at Riverside, Riverside, California}

\icmlcorrespondingauthor{Umberto Casti}{castiumber@dei.unipd.it}
\icmlcorrespondingauthor{Giacomo Baggio}{baggio@dei.unipd.it}
\icmlcorrespondingauthor{Sandro Zampieri}{zampi@dei.unipd.it}
\icmlcorrespondingauthor{Fabio Pasqualetti}{fabiopas@ucr.edu}

\icmlkeywords{Control Theory, Linear SSM, State-Space Models, Induction Head, Sequence Modeling}

\vskip 0.3in
]



\printAffiliationsAndNotice{}  

\begin{abstract}
  A key claim in recent work on Selective State Space Models is that
  selectivity—the ability to focus on relevant information while
  filtering irrelevant inputs—requires breaking the Linear
  Time-Invariant (LTI) property through time-varying dynamics. We
  challenge this claim by demonstrating that LTI systems can achieve
  selectivity when designed using principles from geometric control.

  We introduce the Geometric SSM, in which different input patterns
  excite distinct invariant subspaces of the dynamics. Unlike Mamba's
  memoryless selection mechanism, our approach employs a dynamic
  residual generator that maintains temporal memory, enabling
  recognition of multi-token patterns without time-varying system
  matrices. The Geometric SSM achieves near-perfect performance on a
  novel extended induction head task where Mamba fails, while
  preserving efficient FFT-based training. Our results demonstrate
  that geometric control theory can inform the design of novel
  selective sequence models that combine theoretical rigor with
  practical efficiency.
\end{abstract}

\section{Introduction}\label{sec:intro}
Sequence modeling tasks lie at the core of modern machine learning,
from natural language processing to time-series forecasting. A
fundamental challenge in these applications is efficiently capturing
dependencies and contextual relationships within input
sequences. Recent advances have been driven by two paradigms:
Transformers~\cite{AV-NS-NP-JU-LJ-ANG-LK-IP:17} with attention
mechanisms~\cite{DB-KC-YB:14}, including efficiency-oriented variants \cite{AK-AV-NP-FF:21,KMC-VL-DD-XS-AG-TS-PH-JQD-AM-LK-DBB-LJC-AW:21},  and State Space Models
(SSMs)~\cite{AG-KG-CR:21,JS-AW-SWL:23,AG-TD:23,RH-ML-TW-MC-AA-DR:23,TD-AG:24,NMC-AO-BW-CS-TL:24,LZ-AS-YD-ASG-YS-BB-MT-AA-SS:24}, each offering
distinct trade-offs between expressiveness and computational
efficiency.

Transformers excel at capturing long-range dependencies through
attention but require significant computational resources, scaling
quadratically with sequence length and limiting their applicability to
latency- or memory-constrained
settings~\cite{YT-MD-DB-DM:22}. SSM-based methods offer an alternative
by modeling sequences through linear dynamical systems, achieving
linear complexity while maintaining strong performance. The Mamba
architecture~\cite{AG-TD:23} represents a significant advancement in
this direction, introducing \emph{selective} SSMs that dynamically
modulate which inputs to process based on content. Mamba achieves this
selectivity by making system matrices input-dependent, effectively
introducing Linear Time-Varying dynamics.

A central claim in~\cite{AG-TD:23} is that this time-varying behavior
is \emph{necessary} for selectivity—that Linear Time-Invariant (LTI)
systems fundamentally cannot process inputs selectively. This claim
has important implications: time-varying dynamics break the
convolutional structure of traditional SSMs, sacrificing
parallelization benefits and complicating analysis. If selectivity
indeed requires abandoning LTI properties, this represents an
unavoidable cost.

\noindent
\textbf{Our contribution.} In this work, we demonstrate that this
claim is incorrect. Drawing on geometric control
theory~\cite{GB-GM:91}, we show that LTI systems \emph{can} achieve
selectivity when properly designed. Our key insight is that different
input patterns can be engineered to excite distinct invariant
subspaces of the state space, enabling content-dependent responses
without time-varying system matrices. This geometric perspective,
widely used in fault detection and
isolation~\cite{GM-FM-LN-DM:10,WMW:85}, provides a principled
framework for designing selective LTI architectures.

Building on this insight, we introduce the \emph{Geometric SSM}, an
architecture that moves selection outside the core recurrent
dynamics. Rather than making state matrices input-dependent as in
Mamba, we employ a dynamic \emph{residual generator}—itself an LTI
system—that produces a selection signal based on temporal patterns in
the input sequence. This signal controls a gating mechanism that
interpolates between propagating new information and preserving
historical context, achieving an effect similar to Mamba's
time-varying discretization while maintaining LTI structure.

The Geometric SSM offers several advantages over time-varying
approaches. By preserving LTI properties, we can leverage efficient
input-output (I/O) representations for training, achieving
parallelization through FFT-based convolution without requiring
diagonal system matrices. The architecture provides explicit control
over memory capacity, and the separation of feature extraction,
processing, and selection enhances modularity and interpretability.

\noindent
\textbf{Scope and objectives.} We emphasize that our goal is not to
develop a general-purpose foundation model competitive with full-scale
Mamba implementations. Rather, we focus on understanding the
\emph{mechanisms} that enable selectivity and demonstrating that LTI
architectures suffice for selective tasks when properly designed. Our
experimental evaluation centers on synthetic benchmarks that isolate
selective capabilities, including: (i) the standard Induction Head
task~\cite{AG-TD:23,DYF-TD-KKS-AWT-AR-CR:22}, where a single trigger
token cues recall; (ii) a novel extended induction head task, where
multi-token trigger sequences require temporal memory for recognition;
and (iii) sequential MNIST~\cite{JS-AW-SWL:23}, evaluating performance
beyond purely selective tasks.

The Extended Induction Head task is particularly revealing: it exposes
a fundamental limitation of Mamba's memoryless selection
parametrization, which depends only on the current input without
temporal context. The Geometric SSM, through its dynamic residual
generator, naturally handles such tasks by maintaining memory of past
inputs.


The remainder of this paper is organized as follows: Section~2
analyzes the Mamba architecture and its tradeoffs, Section~3
introduces the Geometric SSM and its theoretical foundations,
Section~4 presents experimental results, and Section~5 concludes with
discussion and future directions.

\section{Selective State Space Models: the Mamba architecture and its
  limitations}\label{sec: selective ssm}
SSMs have emerged as a promising alternative to
Transformers for sequence modeling tasks. While Transformers employ
highly effective attention mechanisms, they are computationally
expensive, relying on large attention matrices that store relational
information and slow down inference. Recurrent Neural Networks (RNNs),
though more efficient in maintaining context through state vectors,
typically underperform compared to Transformers due to inefficient and
complex training processes that are often not
parallelizable~\cite{YT-MD-DB-DM:22}.

The Mamba architecture~\cite{AG-TD:23} represents a significant
advancement in SSMs by introducing a selection mechanism that aims to
combine the advantages of both Transformers and RNNs while mitigating
their respective drawbacks. The key insight behind Mamba is that
\emph{selectivity} is a fundamental property for compression and
comprehension, as it enables the model to focus on specific, relevant
parts of the input while excluding non-informative or irrelevant
components. An effective selection mechanism improves efficiency in
storing information and enhances accuracy in processing inputs, as it
inherently functions as an attention mechanism for extracting useful
information.

\subsection{Architecture overview}
Building on prior works such as S4~\cite{AG-KG-CR:21} and
S5~\cite{JS-AW-SWL:23}, the Mamba architecture introduces selectivity
by replacing a core Linear Time-Invariant (LTI) dynamical system with
a Linear Time-Varying (LTV) dynamical system. Rather than processing
the input $\ut \in \Rm$ through a single Multi-Input Multi-Output
(MIMO) system, Mamba employs a collection of $m$ Single-Input
Single-Output (SISO) systems operating in parallel. Each system
$\Sigma^i_t$, for $i = 1, \ldots, m$, is defined as:
\begin{equation}\label{eq:mamba_system}
  \Sigma^i_t :
  \left\lbrace
    \begin{array}{r@{\;}l}
                \hitOne &= A_t^i \hitime + B^i_t \uit , \\
      \yit &= C_t \hitime,
    \end{array}
  \right.
\end{equation}
where $\uit$ and $\yit$ denote the $i$-th components of the input and
output vectors $\ut$ and $\yt$, respectively. The time-varying system
matrices are defined as:
\begin{subequations} \label{eq:mamba_matrices}
  \begin{align}
    A_t^i &= \exp \big(\Delta^i_t \Ab^i \big), \label{eq:mamba_A}\\
    B^i_t &= \big(\Ab^i\big)^{-1} \big(\exp \big(\Delta^i_t \Ab^i \big) - I \big)\Bb_t, \label{eq:mamba_B}\\
    C_t &= \Cb_t^\top,\label{eq:mamba_C}
\end{align}
\end{subequations}
where $\Ab^i \in \Rnn$ are constant diagonal matrices with negative real eigenvalues (Hurwitz stable), and the time-dependent components are parametrized as:
\begin{subequations}
\label{eq:mamba_parametrization}
\begin{align}
    \Delta^i_t &= \tau\left(W^i_\Delta \ut\right), \label{eq:mamba_delta} \\
    \Bb_t &= W_B \ut, \label{eq:mamba_B_param} \\
    \Cb_t &= W_C \ut, \label{eq:mamba_C_param}
\end{align}
\end{subequations}
where $W^i_\Delta \in \ROnem$, $W_B, W_C \in \Rnm$ are learnable
matrices, and $\tau:\RR \to \RR_+$ is the softplus
function~\cite{SRD-SKS-BBC:22} ensuring $\Delta^i_t >0$ at all times.
Notice that the structure of the matrices $A^i_t$ and $B^i_t$ in
equations~\eqref{eq:mamba_A}--\eqref{eq:mamba_B} directly corresponds
to a Zero-Order Hold (ZOH) discretization~\cite{CLP-HTN:98} of a
continuous-time LTV system with a time-varying sampling interval
$\Delta^i_t$.

\subsection{The selection mechanism}
The core innovation of Mamba lies in the time-varying sampling time
$\Delta_t^i$, which modulates input influence. When
$\Delta_t \to +\infty$, we have $A_t \to 0$ in~\eqref{eq:mamba_A} and
the system attends strongly to the current input $\ut$, while when
$\Delta_t \to 0$, we have $A_t \to I$ and $B_t \to 0$, causing the
system to ignore $\ut$ and preserve the previous state $\htime$. This
dynamic gating allows Mamba to selectively focus on relevant inputs
while filtering less significant ones, implementing content-based
reasoning without quadratic complexity. While additional
time-dependent components appear in equations~\eqref{eq:mamba_B_param}
and~\eqref{eq:mamba_C_param}, they serve primarily as fine-tuning
mechanisms~\cite{AG-TD:23}. Fig.~\ref{fig:wholeMamba} depicts the time-varying selection mechanism.

\begin{figure}[ht]
\vskip -0.5cm
\begin{center}
\centerline{\IfFileExists{./figures/systemConnectionMamba.tex}{\pgfdeclarelayer{background}
\pgfdeclarelayer{foreground}
\pgfsetlayers{background,main,foreground}
\begin{tikzpicture}[node distance=0mm and 5mm, every node/.style=draw, every join/.style = {-latex}]
{[start chain=input going below]
	\node [blank, on chain,label={above left:$\ut$}](input1) {};
	{ [start branch=Sigma1 going right] 
	\node [block, on chain,join,join = with input-1 by {myRed,to path = {(\tikztostart.north) --($(\tikztostart.north) + (0,0.5)$)  -| node[draw = none,midway,above,myRed]{selective action} (\tikztotarget.north)}} ](S1) {$\Sigma^1_t$};
	
	\node [blank, on chain,join,label={above:$\yt$}](output1) {};
	{ [start branch=ouput going below]
	\node [blank, on chain] {};
	\node [blank, on chain] {}; 
	\node [blank, on chain](ouputm){}; 
	} 
	} 
	
	\node [blank, on chain] {};
	{ [start branch= tmp going right] } 
	
	\node [blank, on chain] {};
	{ [start branch= tmp going right] }
	
	\node [blank, on chain](inputm) {};
	{ [start branch=Sigmam going right] 
	\begin{pgfonlayer}{background}
	\node [block, on chain,join, join= with ouputm by {latex-},join = with input-1 by {myRed,to path = {[-](\tikztostart.north) --($(\tikztostart.north) + (0,0.5)$) -| ($(S1.north) + (-.25,0.0)$);\draw[myRed,-] ($(S1.south) + (-.25,0.0)$) -- ($(S1.south) + (-.25,-0.25)$) ($(\tikztotarget.north) + (-.25,0.25)$) [-latex]-- ($(\tikztotarget.north) + (-.25,0.0)$);\draw[myRed,dotted] ($(S1.south) + (-.25,-0.3)$) -- ($(\tikztotarget.north) + (-.25,0.25)$)}}](Sm) {$\Sigma^m_t$};
	\end{pgfonlayer}
	{[start branch=connectSigma going above]
	\chainin (S1) [join = by {dotted,-,shorten >=2pt,shorten <=3pt}]; }
	}
}           
\end{tikzpicture}}{?}}
\caption{Time-varying selection mechanism in Mamba.}
\label{fig:wholeMamba}
\end{center}
\vskip -0.5cm
\end{figure}

\subsection{Drawbacks of time-varying dynamics}
The introduction of time-varying dynamics through input-dependent
parametrization represents both the strength and limitation of the
Mamba architecture. By making the system matrices functions of the
input sequence, Mamba achieves selectivity that rivals
Transformer-based attention mechanisms. However, this design choice
comes at a cost: the time-varying nature disrupts the convolutional
perspective inherent in earlier SSM architectures like S4 and S5.

In LTI SSMs, the relationship between input and output can be
expressed as a convolution with a fixed impulse response, enabling
efficient parallel computation through Fast Fourier Transform (FFT)
algorithms. The time-varying parametrization in Mamba breaks this
property, necessitating sequential computation during training and
inference. While this maintains linear complexity in sequence length
(unlike the quadratic complexity of Transformers), it sacrifices some
of the parallelization benefits of purely convolutional models.

Nevertheless, Mamba demonstrates that breaking time-invariance in a
structured, input-dependent manner provides a powerful mechanism for
selective information processing in sequence modeling tasks, achieving
state-of-the-art performance on various benchmarks while maintaining
computational efficiency. A key claim in~\cite{AG-TD:23} is that the
additional computational cost induced by time-varying architectures is
unavoidable, as LTI architectures are fundamentally unable to process
inputs selectively. The main contribution of this paper is to
demonstrate that this claim is incorrect, showing that LTI
architectures can process inputs selectively and solve complex
selective tasks efficiently.

\section{Geometric SSM}\label{sec: geometric ssm}

\subsection{Selectivity in LTI systems: a geometric perspective}
We now present an example demonstrating that LTI systems can process
inputs selectively, contrary to claims in~\cite{AG-TD:23}. Our
approach leverages geometric control theory~\cite{GB-GM:91}, a
well-established framework for solving structural control problems
including input decoupling, fault detection and isolation, and model
tracking~\cite{GM-FM-LN-DM:10,WMW:85}.

As is common practice, we assume the model operates in Euclidean space, with input sequences properly mapped to vector sequences through embedding~\cite{TM-KC-GSC-JD:13,TM-IS-KC-GC-JD:13}. Consider a simplified \emph{selective copying} task involving two token types: a data token that should be recognized and memorized, and a blank token that should be discarded. We represent these tokens as vectors in $\mathbb{R}^3$:
\begin{align}\label{eq: embedding}
  \text{data-vector} = 
  \begin{bmatrix}
    -4.11  \\ 4.58 \\ 0.60
  \end{bmatrix}
  , \quad
  \text{blank-vector} = 
  \begin{bmatrix}
    9.05 \\ -11.34 \\ -0.04
  \end{bmatrix}.
\end{align}

We construct a sequence alternating randomly between blank and data
vectors, and feed this sequence to the following discrete-time LTI
system:
\begin{align}\label{eq: example system}
  A &=
      \begin{bmatrix}
        1.48    & 1.14   & 0.42\\
        -1.36  &-1.04  & -0.16\\
        0.01    & 0.01   & 0.46
      \end{bmatrix}, \quad
  B =
    \begin{bmatrix}
      1   & 0   & 0\\
      0  & 1  & 0\\
      0    & 0   & 1
    \end{bmatrix},\nonumber \\
  C^\top &=
           \begin{bmatrix}
             0.1270  \\ 0.0975  \\ 0.9575
           \end{bmatrix}.
\end{align}

\begin{figure}[t]
    \centering
    \includegraphics[width=1\columnwidth]{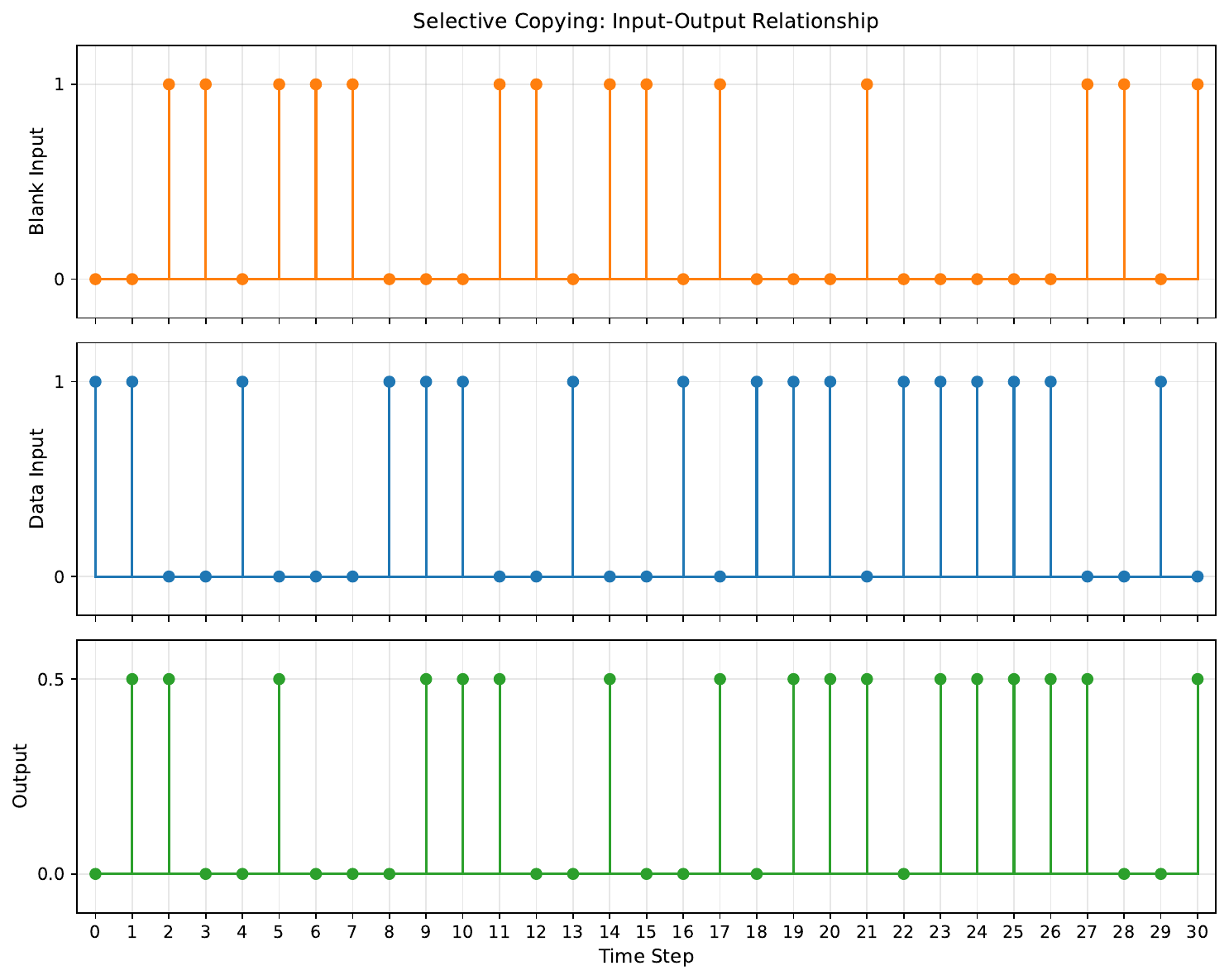}
    \caption{Selective copying with an LTI system: the system
      correctly filters out blank vectors (producing zero output)
      while generating a non-zero response to data vectors. The output
      at time $t$ equals $0.5$ when the input at time $t-1$ is a data
      vector, and zero otherwise. Numerical values in \eqref{eq:
        embedding} and \eqref{eq: example system} are rounded for clarity.}
    \label{fig: selective_copying}
\end{figure}

The input-output relationship is illustrated in Figure~\ref{fig:
  selective_copying}. For instance, at times $t = 0, 1$, the data
vector is fed into the system, while at times $t = 2, 3$, the blank
vector is provided. Remarkably, the LTI system~\eqref{eq: example
  system} correctly filters out the blank vectors while allowing data
vectors to generate a non-zero output signal. Specifically, the system
output at time $t$ equals $0.5$ when the input at time $t-1$ is a data
vector, and zero otherwise. In other words, the LTI system~\eqref{eq:
  example system} processes inputs \emph{selectively}, generating a
non-zero response only for data vectors. This demonstrates that
carefully designed discrete-time LTI systems are capable of
content-dependent responses that can be exploited for selective
computation.

\noindent
\textbf{Geometric interpretation and generality.}  The mechanism
underlying this example is not a pathological case but rather reflects
fundamental principles from geometric control
theory~\cite{GB-GM:91}. The key insight is that different input
vectors can be designed to excite distinct invariant subspaces of the
state space, enabling the system to produce differential responses
based on input content. This geometric perspective naturally extends
to handle multiple distinct input classes beyond binary data/blank
categorization, tailored dynamic responses for different input types,
more sophisticated selection processes where responses to different
vectors are confined to non-intersecting subspaces, as well as noisy
and approximate settings. Furthermore, this geometric approach can be
combined with static gating mechanisms and LTI-based memory components
to construct purely LTI architectures with capabilities similar to
modern selective SSMs, while maintaining favorable training properties
and computational complexity.

Inspired by this geometric perspective, our proposed architecture
moves the selection mechanism \emph{outside} the core recurrent
relationship, in contrast to Mamba's approach of introducing
time-varying dynamics within the recurrence itself
(Equations~\eqref{eq:mamba_system}--\eqref{eq:mamba_matrices}). This
design choice enables us to fully leverage the advantageous properties
of LTI systems during
training~\cite{NCR-CE:21,RNP-SM-AM-JTHS-RH-ML-QA-CR-HA-SE-TS-MP-AY:24}
while avoiding restrictive structural constraints such as the
requirement for diagonal system matrices.

\subsection{Description of the proposed architecture}
Our proposed architecture is illustrated in
Figure~\ref{fig:ctrlScheme}. The design involves three LTI
systems---$\Sigma_f$, $\Sigma_M$, and $\Sigma_r$---combined with a
nonlinear gating mechanism $\Sigma_g$. The information flow is as
follows:

\begin{enumerate}[topsep=0pt,itemsep=0pt]

\item The input $\ut$ is processed by $\Sigma_f$ to generate a
  ``signature'' $\ft$ that captures salient features of the input.
    
\item Both $\ut$ and $\ft$ are fed into $\Sigma_M$, which produces a
  candidate output $\yst$.
    
\item The residual system $\Sigma_r$ computes a residual signal $\rt$
  based on the difference between $\yst$ and $\ut$, generating the
  selection signal $\st = \sigma(\rt)$ through a sigmoid activation.
    
\item Finally, the gating mechanism $\Sigma_g$ uses $\st$ to
  interpolate between the previous output $\yt$ and the candidate
  output $\yst$, producing the final output $\ytOne$.

\end{enumerate}

\begin{figure}[ht]
  \vskip -0.1em
  \begin{center}
    \centerline{\IfFileExists{./figures/myFigBlockScheme2.tex}{\begin{tikzpicture}[ every join/.style = {-latex},node distance=2mm and 1cm]
{ [start chain=trunk]
    \node [input,on chain,label = {[shift={(-0.45,0)}]above right:$\ut$}] (input) {};
    \coordinate[on chain = going {at=(\tikzchainprevious),shift=(0:0.25)}, join = by -] (sep);
    
    {[start branch = faultBranch going {at=(\tikzchainprevious),shift={(0.75,0.25)}}]
    \node[blockSmall,on chain,join = with trunk-2 by {to path ={(\tikztostart) |- (\tikztotarget)}}] (sysFault) {$\Sf$};
    }
    
    { [start branch = uBranch going {at=(\tikzchainprevious),shift={(1,-0.25)}}]
    \coordinate[on chain,join = with trunk-2 by {-,to path ={(\tikztostart) |- (\tikztotarget)}}] (sysFaultb);
  	}
  	
  	\node [blockSmall,on chain = going {at=(\tikzchainprevious),shift=(0:1.1)} ,join= with trunk/faultBranch-end by {to path = {(\tikztostart) -- node[above] {$\ft$} (\tikztostart-|\tikztotarget.west)}},join= with trunk/uBranch-end by {to path = {(\tikztostart) -- (\tikztostart-|\tikztotarget.west)}},minimum size=0.75cm] (sys) {$\Sigma_M$};
  	\coordinate [on chain = going {at=(\tikzchainprevious),shift=(0:0.25)},join = by -] (sysResGena) {};
  	
  	{[start branch = resGenBranch going {at=(\tikzchainprevious),shift={(0.75,-1.0)}}]
\node[blockSmall,on chain, join = with sysResGena by {to path ={(\tikztostart) |- ($(\tikztotarget.west) + (0,+0.125)$)}} ,join = with sysFaultb by {to path ={(\tikztostart) |- ($(\tikztotarget.west) + (0,-0.125)$)}}] (resGen) {$\Sigma_r$};

\node[sigmoid,on chain = going {at=(\tikzchainprevious),shift={(0.5,0)}}, join = by {label = { [shift  = {(0.35,0.1)}] right :$\st$}}] (decMaker) {};}
  	
  	\node [blockSmall,text  = myBlue, draw = myBlue,on chain = going {at=(\tikzchainprevious),shift=(0:1.2)}, join = by {to path ={(\tikztostart) --  node[above] {$\yst$} (\tikztotarget)}},join = with decMaker by {to path ={(\tikztostart) -| (\tikztotarget)}} ]  (selection) {$\Sigma_g$};
  	
  	\node [output,on chain = going {at=(\tikzchainprevious),shift=(0:0.45)}, join,label = {[shift={(0:0.2)}]above left:$\yt$}] (output) {};

\node[fit=(sep)(sysFault)(sysFaultb)(sys),inner ysep=0.30cm,inner xsep=0.20cm, draw,densely dotted, gray,  label = {[gray,above right]south west:$\Sigma$ LTI}] {};

\node[fit=(resGen)(decMaker),inner xsep=0.20cm, draw,densely dotted, myRed,  label = {[myRed]south:residual selection}] {};

}

\end{tikzpicture}}{?}}
    \caption{Block-diagram representation of the Geometric SSM. The
      systems $\Sigma_f$, $\Sigma_M$, and $\Sigma_r$ are LTI, while
      $\Sigma_g$ represents a nonlinear gating mechanism. The
      selection signal $\st$ acts as an attention mechanism that
      controls the propagation of information.}
\label{fig:ctrlScheme}
\end{center}
\vskip -0.6em
\end{figure}

While this structure may appear complex, it is designed to be
efficiently learnable: all trainable parameters are embedded within
LTI systems, while nonlinearities remain either static (the sigmoid
function) or structurally fixed (the gating operation). During
training, a single LTI system composed of $\Sigma_f$ and $\Sigma_M$ is
learned, since parallel and series interconnections of LTI systems are
guaranteed to generate an LTI system~\cite{JPH:18}. The separation
into $\Sigma_f$ and $\Sigma_M$ is presented here only for
interpretability reasons.

The proposed architecture offers several advantages over purely
time-varying approaches:

\noindent
\textbf{Preserved LTI structure:} all trainable system matrices belong
to LTI systems, enabling efficient training algorithms and theoretical
analysis tools developed for LTI SSMs.

\noindent
\textbf{Flexible parameterization:} the architecture does not
require diagonal matrices or other structural constraints, allowing
for richer representational capacity.

\noindent
\textbf{Modular design:} the separation of feature extraction
($\Sigma_f$), processing ($\Sigma_M$), residual computation
($\Sigma_r$), and gating ($\Sigma_g$) provides interpretability and
modularity.

\noindent
\textbf{Dynamic selection:} the signal $\st$ depends on the residual
between the candidate output and input, providing a content-based
attention mechanism that adapts to sequence context.

This design shifts complexity from recurrence-based nonlinearities to
a structured combination of LTI systems with static nonlinear
components, maintaining computational efficiency while enabling
selective information processing.

\subsection{Mathematical formulation}
We now provide the precise mathematical definitions of each component
in the proposed architecture.

\noindent
\textbf{Signature system $\Sigma_f$:}
\begin{equation}
\label{eq:sigma_f}
\Sigma_f : \left\lbrace\begin{array}{r@{\;}l}
\hftOne &= A_f \hftime + B_f \ut\\ 
\ft &= C_f \hftime + D_f \ut 
\end{array}\right.,
\end{equation}
where $A_f \in \Rnfnf$, $B_f \in \Rnfm$, $C_f \in \Rmnf$, and
$D_f \in \Rmm$ are learnable parameters.

\noindent
\textbf{Main processing system $\Sigma_M$:}
\begin{equation}
\label{eq:sigma_M}
\Sigma_M : \left\lbrace\begin{array}{r@{\;}l}
\hMtOne &= A_M \hMtime + B_M^\prime \ut + B_M^{\prime\prime} \ft \\ 
\yst &= C_M \hMtime + D_M^\prime \ut + D_M^{\prime\prime} \ft
\end{array}\right.,
\end{equation}
where $A_M \in \RnMnM$, $B_M^\prime, B_M^{\prime\prime} \in \RnMm$, $C_M \in \RmnM$, and $D_M^\prime, D_M^{\prime\prime} \in \Rmm$ are learnable parameters.

\noindent
\textbf{Residual system $\Sigma_r$:}
\begin{equation}
\label{eq:sigma_r}
\Sigma_r : \left\lbrace \begin{array}{r@{\;}l}
\hrtOne &= A_r \hrtime + B_r \left(\yst - \ut \right) \\ 
\rt &= C_r \hrtime + D_r \left(\yst - \ut \right)
\end{array}\right.,
\end{equation}
where $A_r \in \RnResnRes$, $B_r \in \RnResm$, $C_r \in \ROnenRes$,
and $D_r \in \ROnem$ are learnable parameters. The residual
$\yst - \ut$ serves as the input to this system, analogous to
residuals used in fault detection.

\noindent
\textbf{Gating mechanism $\Sigma_g$:}
\begin{equation}
\label{eq:gating}
\Sigma_g \,:\, \ytOne = \yt + \left(\yst - \yt \right)\st,
\end{equation}
where $\st \coloneqq \sigma\left(\rt\right)$ and
$\sigma:\RR \to (0, 1)$ denotes the sigmoid
function~\cite{SRD-SKS-BBC:22}.

The gating equation~\eqref{eq:gating} implements a convex combination:
when $\st \approx 1$, the system propagates the new candidate output
$\yst$, effectively attending to current information; when
$\st \approx 0$, the system retains the previous output $\yt$,
preserving historical context and ignoring the current input.

\noindent
\textbf{Model parameterization.} The state dimensions are defined as
$n_f = m\nu_f$, $n_M = m\nu_M$, and $n_r = \nu_r$, where $\nu_f$,
$\nu_M$, and $\nu_r$ are positive integers controlling the memory
capacity of each subsystem. These parameters determine the temporal
extent over which the systems can integrate past information,
providing explicit architectural control over long-range dependency
modeling without requiring time-varying dynamics.

\subsection{Efficient implementation via I/O
  representations}\label{sec:implementation}
A key feature of Mamba and related selective SSMs is their
parallelizable training procedures that maintain moderate
computational and memory requirements. This is achieved through
parallel scan algorithms~\cite{GEB:90} and by constraining the state
matrix $A$ to be diagonal. Building
on~\cite{RNP-SM-AM-JTHS-RH-ML-QA-CR-HA-SE-TS-MP-AY:24}, we show that
the proposed Geometric SSM achieves similar—and in some respects
superior—efficiency without enforcing structural constraints. Rather
than relying on scan-based algorithms, the Geometric SSM leverages the
I/O representation of LTI systems, a characterization available
exclusively for time-invariant systems. This choice preserves
expressivity while yielding a more compact parameterization and
enabling efficient, parallelizable training through FFT-based
convolution.

\subsubsection{I/O parameterization} 
All learnable parameters of the Geometric SSM, except for the vector
embeddings, are contained within the LTI systems $\Sigma_f$,
$\Sigma_M$, and $\Sigma_r$. Each of these systems admits two
equivalent representations: a state-space representation, defined
through recurrence relations as described in Section~3.3, and an I/O
representation, expressed in terms of its transfer function via the
$\mathcal{Z}$-transform~\cite{JPH:18}. Crucially, the I/O
representation is available only for LTI systems.

The key advantage of this representation is its parameter
efficiency. In the I/O representation, the interconnected system
$\Sigma$ is represented as an $m\times m$ rational matrix-valued
transfer function of order $n_\Sigma = n_f + n_M$, while the residual
generator $\Sigma_r$ is represented as a $1\times m$ rational transfer
function of order $n_r$. In this formulation, the learnable parameters
consist of the coefficients of the numerator and denominator
polynomials of the transfer functions, together with direct
feedthrough terms. For $\Sigma$, the I/O representation requires
$m^2 n_\Sigma$ parameters for the matrix-valued numerator
coefficients, $n_\Sigma$ parameters for the denominator, and $m^2$
parameters for the feedthrough matrix. For $\Sigma_r$, $m n_r$
numerator coefficients, $n_r$ denominator coefficients, and $m$
feedthrough parameters are required. In summary, the total number of
learnable parameters scales linearly with $n_\Sigma$ and $n_r$—recall
from Section~3.3 that these parameters govern the model's memory
capacity through $\nu_f$, $\nu_M$, and $\nu_r$.

We emphasize that a state-space parameterization of the same
architecture would require storing the system matrices
$A_\Sigma \in \mathbb{R}^{n_\Sigma \times n_\Sigma}$ and
$A_r \in \mathbb{R}^{n_r \times n_r}$. Without structural constraints,
this leads to a quadratic dependence on state
dimension, namely $\mathcal{O}(n_\Sigma^2 + n_r^2)$ parameters for the state
matrices alone. The I/O representation eliminates this quadratic
scaling, requiring a number of parameters that scales
linearly with $n_\Sigma$ and $n_r$ while
preserving the expressivity of a dense state matrix.

For comparison, Mamba retains a state-space parameterization but
constrains the state matrix $A$ to be diagonal. For a model with state
dimension $n$, this yields $mn$ parameters for the state dynamics,
with additional $2mn$ parameters arising from input-output projections
$W_B$ and $W_C$, and $m^2$ parameters from input-dependent
discretization (matrices $W_{\Delta}^i$). Thus, as in Geometric SSMs,
the number of parameters in Mamba scales linearly with the state
dimension $n$. 

\subsubsection{Parallelizable training}
Beyond parameter efficiency, the I/O representation enables favorable
training-time computation and memory scaling.

In the I/O representation, the model is characterized directly by its
matrix-valued transfer functions, allowing training to proceed in a
state-free manner without explicitly propagating hidden states across
time~\cite{NCR-CE:21,RNP-SM-AM-JTHS-RH-ML-QA-CR-HA-SE-TS-MP-AY:24}. For
an input sequence of length $\ell$, the output sequence is computed
via FFT-based convolution: the input is transformed to the frequency
domain, multiplied element-wise by the transfer functions evaluated on
a truncated Fourier grid, and transformed back to the time
domain. This procedure is fully parallelizable and incurs
$\mathcal{O}(\ell \log \ell)$ cost for FFT operations. For the
Geometric SSM, frequency-domain computation requires multiplying an
$m \times m$ transfer matrix by an $m$-dimensional vector at each of
$\ell$ frequencies, yielding total computational cost
$\mathcal{O}(\ell m^2 + \ell m \log \ell)$. Crucially, no state
trajectories are stored during training—only input and output vectors
indexed by time and channel—leading to memory requirements of
$\mathcal{O}(\ell m)$, independent of the dimensions $n_{\Sigma}$ and~$n_r$.


By contrast, Mamba enables parallelizable training via scan-based execution of time-varying state-space recursions~\cite{GEB:90}. While this provides logarithmic parallel depth in the sequence length, the procedure remains state-based and requires $\mathcal{O}(\ell m n)$ memory to store intermediate states for all SISO subsystems. The diagonal constraint on $A$ reduces per-step computation to $\mathcal{O}(n)$ per subsystem; accounting for $m$ subsystems and input-dependent parameter generation, this yields a total training cost of $\mathcal{O}(\ell(mn + m^2))$.

We summarize parameter scaling and training complexity for Geometric
SSMs and Mamba in Table~\ref{tab:geometric-vs-mamba}.
\begin{table}[h]
\centering
\setlength{\tabcolsep}{3pt}
\renewcommand{\arraystretch}{1.15}
\small
\begin{tabular}{lcc}
\toprule
\textbf{Model} & \textbf{Parameters} & \textbf{Training (Comp. / Memory)} \\
\midrule
Geom.~SSM 
& $\mathcal{O}(m^2 n_\Sigma\! +\! m n_r)$ 
& $\mathcal{O}(\ell (m^2\! +\! m \log\ell))$ / $\mathcal{O}(\ell m)$ \\
Mamba
& $\mathcal{O}(mn\! +\! m^2)$ 
& $\mathcal{O}(\ell (mn+m^2))$ / $\mathcal{O}(\ell m n)$ \\
\bottomrule
\end{tabular}
\caption{Parameter scaling and training complexity. Here $\ell$ is
  sequence length, $m$ is embedding dimension, $n_{\Sigma}=n_f+n_M$,
  and $n$ is the state dimension of each SISO subsystem in Mamba.}
\label{tab:geometric-vs-mamba}
\end{table}

The comparison reveals complementary efficiency profiles. Geometric
SSMs incur quadratic cost in the embedding dimension $m$ but scale
independently of the internal state dimensions during training, while
Mamba’s computational cost scales with the state dimension $n$ and the
number of subsystems $m$, reflecting its state-based computation.
Accordingly, Geometric SSMs are favored when large state dimensions are
required, whereas Mamba can be more efficient when the per-subsystem
state dimension is small. In practice, both models achieve efficient
training through parallelization, with relative performance depending
on the problem-specific values of $m$, $n$, and $\ell$.


\section{Experimental evaluation}\label{sec:experiments}
We now evaluate the Geometric SSM architecture and compare it against
the Mamba selective mechanism on carefully designed synthetic
tasks. Our primary objective is to investigate the fundamental
properties of different selection mechanisms rather than develop a
complete production-ready architecture. To this end, we isolate the
core Selective SSM component from Mamba, as presented in Algorithm 2
of~\cite{AG-TD:23}, and evaluate both selection mechanisms on the
tasks described in Section~\ref{subSec:synTask}. We focus on their
structural differences and the distinct selection strategies they
implement. Additionally, we include experiments on sequential MNIST
(sMNIST) to evaluate performance on a standard benchmark where
selectivity is less central to success. All experiments are conducted
on hardware equipped with an NVIDIA A40 GPU with 48 GB of memory.

\subsection{Motivating tasks}\label{subSec:synTask}
We consider two synthetic tasks that illustrate fundamental challenges
in selective sequence modeling and in-context
learning~\cite{CO-NE-NN-NJ-ND-TH-BM-AA-YB-AC-TC-DD-DG-ZHD-DH-SJ-AJ-JK-LL-KN-DA-TB-JC-JK-SM-CO:22,JC-ES:24}. While
relatively simple, these tasks highlight key differences between
static and dynamic selection mechanisms.

\noindent
\textbf{Induction head.} The induction head task, introduced
in~\cite{AG-TD:23,DYF-TD-KKS-AWT-AR-CR:22}, represents a context-aware
mechanism for in-context
learning~\cite{CO-NE-NN-NJ-ND-TH-BM-AA-YB-AC-TC-DD-DG-ZHD-DH-SJ-AJ-JK-LL-KN-DA-TB-JC-JK-SM-CO:22}. The
task requires the model to memorize information following a trigger
token that appears exactly twice in a sequence: once in the middle and
once at the end. The objective is to recall the token that followed
the first occurrence of the trigger after observing the second
occurrence. This tests the model's ability to selectively attend to
and retrieve relevant context based on a single-token cue.

\noindent
\textbf{Extended induction head.} We introduce an extension of the
standard induction head task where the trigger is not a single token
but a \emph{sequence} of tokens of length $N_{\text{trig}}$. The model
must recognize this multi-token pattern and recall the token that
followed its first occurrence. This variant fundamentally differs from
the standard task because the selection mechanism must maintain memory
of multiple consecutive tokens to identify the trigger sequence.

Both tasks are illustrated in Figure~\ref{fig:tasks}.

\begin{figure}[t]
\vskip 0.2in
\begin{center}
\centerline{\IfFileExists{./figures/synTasksFig.tex}{\begin{tikzpicture}[node distance=4mm and 0mm,
                    every node/.style=draw]
{ [start chain=trunk]
    \node [token,draw= myBlue,fill=myBlue,on chain] (startt) {};
    { [start branch = b1 going below]
    \node [blank, on chain,join = by -latex](startb) {};
    
  	}
    \node [token,draw = myYellow,fill=myYellow,on chain,token] {};
    { [start branch = b2 going below]
    \node [blank, on chain,join = by -latex] {};
  	}
    \node [trigger,on chain] (triggert) {};
    { [start branch = b3 going below]
    \node [blank, on chain,join = by -latex] {};
    
  	}
    \node [token,draw = myRed,fill=myRed,on chain] {};
    { [start branch = b4 going below]
    \node [blank, on chain,join = by -latex] {};
    
  	}
    
    \node [token,draw = myGreen,fill=myGreen,on chain]  {};
	{ [start branch = b9 going below]
    \node [blank, on chain,join = by -latex] {};
    
  	}
  	\node [trigger,on chain]  {};
	{ [start branch = b9 going below]
    \node [blank, on chain,join = by -latex] {};
    
  	}
  	\node [questionMark,on chain] (endt) {?};
	{ [start branch = b9 going below]
    \node [token, draw = myRed, fill=myRed, on chain,join = by -latex] (endb) {};
    
  	}
}

\draw [braced] (endb.south west) -- node[below left, xshift = 2.5mm, draw = none] {target} (endb.south east) ;
\draw [braceu] (triggert.north west) -- node[above, draw = none] {trigger} (triggert.north east);
{ [start chain=trunk2]
    \node [token,draw = myGreen,fill=myGreen,on chain, right= 2mm of endt] (startt2) {};
    { [start branch = b1 going below]
    \node [blank, on chain,join = by -latex](startb2) {};
    
  	}
    \node [token,draw = myYellow,fill=myYellow,on chain,token](starttriggert2){};
    { [start branch = b2 going below]
    \node [blank, on chain,join = by -latex] {};
  	}
   
    \node [token,draw = myPurple, fill=myPurple,on chain] (endtriggert2) {};
    { [start branch = b4 going below]
    \node [blank, on chain,join = by -latex] {};
    
  	}
    \node [token,draw = myBlue, fill=myBlue,on chain] {};
    { [start branch = b5 going below]
    \node [blank, on chain,join = by -latex] {};
    
  	}
    
    \node [token,draw = myYellow, fill=myYellow,on chain] {};
    { [start branch = b7 going below]
    \node [blank, on chain,join = by -latex] {};
    
  	}
    \node [token,draw = myPurple, fill=myPurple,on chain] {};
    { [start branch = b3 going below]
    \node [blank, on chain,join = by -latex] {};
    
  	}
  	
  	\node [questionMark,on chain] (endt2) {?};
	{ [start branch = b9 going below]
    \node [token,draw = myBlue, fill=myBlue, on chain,join = by -latex] (endb2) {};
    
  	}
  
}

\draw [braced] (endb2.south west) -- node[below left, xshift = 2.5mm, draw = none] {target} (endb2.south east) ;
\draw [braceu] (starttriggert2.north west) -- node[above, draw = none] {trigger} (endtriggert2.north east);
\node [draw = none, above = 7.5mm] at ($(startt)!0.5!(endt)$) {Induction head};
\node [draw = none, above = 7.5mm] at ($(startt2)!0.5!(endt2)$) {Ext. Induction head};
\coordinate (tmpCenterh) at ($(endt)!0.5!(startt2)$);
\coordinate (tmpCenterv) at ($(endt)!0.5!(endb)$);
\coordinate (center) at (tmpCenterh|-tmpCenterv);

\coordinate (tmpTop) at ($(center) + (0,1.75)$);
\coordinate (tmpBottom) at ($(center) - (0,1.25)$);
\draw (tmpTop)  -- (tmpBottom) ;
\end{tikzpicture}}{?}}
\caption{\textbf{Left:} Induction head task, where a single
  trigger token cues recall of the following token. \textbf{Right:}
  Extended induction head task, where a multi-token trigger sequence
  cues recall, requiring a selection mechanism with temporal memory.}
\label{fig:tasks}
\end{center}
\vskip -0.2in
\end{figure}

The extended induction head task poses a particular challenge for
static selection mechanisms. The Mamba selective mechanism
(Equations~\eqref{eq:mamba_parametrization}) depends only on the
current input $\ut$ and does not retain memory of previous
inputs. While the task could theoretically be solved by expanding the
token vocabulary to include all possible trigger sequences of length
$N_{\text{trig}}$, this approach requires a number of embeddings that
grows exponentially with $N_{\text{trig}}$, quickly becoming
impractical. In contrast, the Geometric SSM generates a dynamic
selection signal through the system $\Sigma_r$, which maintains memory
of past inputs. This enables recognition of multi-token patterns
without exponential vocabulary expansion, providing a more scalable
solution.

\subsection{Implementation details}
Both architectures operate on the same vector embedding framework,
using a learnable lookup table for a fixed-size dictionary.

\noindent
\textbf{Mamba Selective SSM.}
We implement the Mamba selective mechanism using
Equations~\eqref{eq:mamba_matrices}
and~\eqref{eq:mamba_parametrization}. For these experiments, we use
recurrent computation during training rather than the selective scan
algorithm. This reflects our focus on conceptual comparison of
selection mechanisms rather than optimized implementation for
large-scale tasks.

\noindent
\textbf{Geometric SSM.}  The Geometric SSM (Section~\ref{sec:
  geometric ssm}) is implemented using the I/O representation
described in Section~\ref{sec:implementation}. During training, we
employ the state-free FFT-based convolution approach
from~\cite{RNP-SM-AM-JTHS-RH-ML-QA-CR-HA-SE-TS-MP-AY:24}, which offers
several advantages: efficient implementation using standard FFT
routines, full parallelization over time, and memory
efficiency independent of internal state dimensions. Following~\cite{RNP-SM-AM-JTHS-RH-ML-QA-CR-HA-SE-TS-MP-AY:24}, we initialize
the model with zero initial conditions, which avoids the need for
explicit stability constraints during training. 
During inference, we employ a recurrent state-space realization,
using the multivariable controllable canonical form
\cite{JPH:18}.

\subsection{Results: induction head}
We replicate the experimental setup
from~\cite{AG-TD:23,TD-DYF-KKS-AWT-AR-CR:23} on sequences of length
$L=16$ with a dictionary of $N=8$ tokens. The Mamba Selective SSM uses
state dimension $n=8$ and embedding dimension $m=16$. The Geometric
SSM uses $\nu_f + \nu_M = 4$, $\nu_r = 4$, and $m=2$. These
configurations balance model simplicity with parameter counts:
approximately $700$ parameters for Selective SSM and $50$ parameters
for Geometric SSM (our model requires fewer parameters due to careful
design).

Following~\cite{AG-TD:23}, we evaluate generalization by testing on sequences of increasing length. Results are shown in Table~\ref{tab:IH}.

\begin{table}[h]
\vskip -.5em
\centering
\small
\setlength{\tabcolsep}{3pt}
\begin{tabular}{lccccccc}
\toprule
\multirow{2}{*}{Model} & \multicolumn{7}{c}{Sequence Length} \\
\cmidrule(lr){2-8}
 & $2^4$ & $2^5$ & $2^6$ & $2^7$ & $2^8$ & $2^9$ & $2^{10}$ \\
\midrule
Sel. SSM & 0.70 & 0.60 & 0.53 & 0.34 & 0.25 & 0.21 & 0.20 \\
Geom. SSM & 0.99+ & 0.99+ & 0.99+ & 0.99+ & 0.99+ & 0.99+ & 0.99+ \\
\bottomrule
\end{tabular}
\vskip -0.5em
\caption{Induction head, validation accuracy. Both models are trained
  on sequences of length $2^4 = 16$.}
  \label{tab:IH}
\end{table}

The Geometric SSM achieves near-perfect accuracy across all sequence
lengths, demonstrating strong generalization. The Selective SSM shows
more limited performance, which appears to contradict results
in~\cite{AG-TD:23} where Mamba achieved perfect accuracy. However, the
work in~\cite{AG-TD:23} uses a two-layer model with state dimension
$n=64$ and approximately 74K parameters, whereas our single-layer
models use substantially fewer parameters (700 and 50,
respectively). This suggests that the original Mamba results may rely
more on model capacity and optimization rather than its selection
mechanisms.

\subsection{Results: extended induction head}
Models are trained on sequences of length $L=16$ with dictionary size
$N=8$ and trigger sequence length $N_{\text{trig}}=4$. We use the same
parameters as in the previous experiment.

\begin{table}[h]
\vskip -0.5em
\centering
\small
\setlength{\tabcolsep}{3pt}
\begin{tabular}{lccccccc}
\toprule
\multirow{2}{*}{Model} & \multicolumn{7}{c}{Sequence Length} \\
\cmidrule(lr){2-8}
 & $2^4$ & $2^5$ & $2^6$ & $2^7$ & $2^8$ & $2^9$ & $2^{10}$ \\
\midrule
Sel. SSM & 0.69 & 0.57 & 0.22 & 0.20 & 0.17 & 0.15 & 0.13 \\
Geom. SSM & 0.99+ & 0.99+ & 0.99+ & 0.99+ & 0.99+ & 0.99+ & 0.99+ \\
\bottomrule
\end{tabular}
\vskip -0.5em
  \caption{Extended induction head, validation accuracy. Both models
    are trained on sequences of length $2^4 = 16$.}
    \label{tab:EIH}
\end{table}

Results are shown in Table~\ref{tab:EIH}. The Geometric SSM maintains
near-perfect performance, generalizing successfully to sequences much
longer than those seen during training. In contrast, the Selective SSM
fails to learn the task effectively.

We attribute this failure to the static nature of the Mamba selection
mechanism (Equations~\eqref{eq:mamba_parametrization}). The selection
parameters $\Delta_t$, $\Bb_t$, and $\Cb_t$ depend only on the current
input $\ut$ without retaining memory of previous tokens. This
memoryless mapping cannot detect trigger sequences that span multiple
time steps at $t-1$, $t-2$, etc.

The Geometric SSM addresses this limitation through the residual
generator $\Sigma_r$, which is itself a dynamical system. This enables
$\Sigma_r$ to recognize multi-token patterns and adapt the selection
signal $\st$ based on temporal context. As discussed in Section~3.3,
the memory parameter $\nu_r$ explicitly controls the temporal extent
over which the residual system integrates information, naturally
accommodating trigger sequences of varying lengths.

\subsection{Results: sequential MNIST}
To evaluate whether the Geometric SSM is useful beyond selection-based
tasks, we test it on sequential MNIST (sMNIST)~\cite{JS-AW-SWL:23}. In
this benchmark, the model receives MNIST image pixels sequentially
(one pixel at a time in raster order) without explicit spatial
structure, creating a long-range dependency problem over 784 time
steps.

We use embedding dimension $m=16$ for both models, with pixel values
from a 256-token vocabulary (grayscale). For Selective SSM, we set
state dimension $n=128$. For Geometric SSM, we use
$n_f = n_M = n_r = 64$.

\begin{table}[h]
  \vspace{-.5em}
\centering
\small
\begin{tabular}{lc}
\toprule
Model & Test Accuracy \\
\midrule
Sel. SSM & 0.11 \\
Geom. SSM  & 0.81 \\
\bottomrule
\end{tabular}
  \vspace{-.5em}
\caption{Test accuracy on sequential MNIST.}
\label{tab:sMNIST}
\end{table}

Results are shown in Table~\ref{tab:sMNIST}. The Geometric SSM
achieves substantially better performance (81\% vs. 11\% accuracy). We
note that the Selective SSM, which relies on time-varying dynamics and
requires storing state trajectories during training, is substantially
more memory-intensive than the Geometric SSM. On our hardware,
increasing the Selective SSM state dimension further results in
prohibitive memory consumption. In contrast, the Geometric SSM's I/O
representation enables scaling dimensions with comparatively modest
memory overhead, as discussed in Section~\ref{sec:implementation}.

While sMNIST is a relatively simple benchmark by modern standards,
these results suggest that the proposed residual-based selection mechanism
can perform well on general sequence modeling problems where explicit
selectivity is not the primary challenge. We view this as an initial
indication that our architecture is not restricted to synthetic
selectivity benchmarks but can transfer to standard long-range tasks.

\subsection{Discussion}
The synthetic tasks, while simple, serve as valuable diagnostic
tools. They reveal fundamental differences between static and dynamic
selection mechanisms: the Mamba approach excels when selection depends
only on individual tokens, but struggles when selection requires
temporal patterns. The extended induction head task specifically
highlights this limitation, demonstrating that the memoryless
parametrization in Equations~\eqref{eq:mamba_parametrization} cannot
capture multi-step dependencies without exponential vocabulary
expansion.

The Geometric SSM's strong performance on both synthetic tasks
validates the core principle that LTI systems can implement effective
selection mechanisms when combined with dynamic residual-based
gating. The sMNIST results, though preliminary, suggest that this approach
extends beyond purely selective tasks to general sequence modeling.

\section{Conclusions}\label{sec:conclusions}
This paper challenges a fundamental assumption in recent selective
State Space Models: that selectivity requires time-varying
dynamics. Using geometric control theory, we have demonstrated that
LTI systems can process inputs selectively when designed to exploit
invariant subspace structures. The proposed Geometric SSM architecture
validates this principle, achieving strong performance on selective
tasks while preserving the beneficial properties of LTI systems.

Our experimental results reveal key insights about selection
mechanisms. On the standard induction head task, the Geometric SSM
achieves near-perfect accuracy with only $50$ parameters, while the
Selective SSM's performance degrades substantially as sequence length
increases despite using $700$ parameters. The extended induction head
task—where trigger sequences span multiple tokens—further exposes the
fundamental limitation of Mamba's memoryless selection
parametrization, with accuracy dropping below $20\%$. By incorporating
temporal memory through the residual generator $\Sigma_r$, the
Geometric SSM naturally handles such multi-step dependencies without
exponential vocabulary expansion, maintaining $99\%+$ accuracy across
all sequence lengths. The sequential MNIST results demonstrate that
the Geometric SSM architecture transfers beyond purely selective tasks
to general sequence modeling, achieving $81\%$ accuracy.


From a practical perspective, the Geometric SSM offers implementation
advantages. The I/O representation enables efficient parallelizable 
training via FFT-based convolution, with memory requirements independent 
of internal state dimensions. Unlike Mamba's diagonal constraints and 
scan-based algorithms, the Geometric SSM supports dense system matrices 
while maintaining computational efficiency.


\noindent
\textbf{Limitations and future work.} Our experiments focus on
relatively simple benchmarks designed to isolate selective
capabilities rather than evaluate complete foundation models. While
the Geometric SSM demonstrates strong performance on these tasks,
scaling to language modeling or other large-scale applications remains
unexplored. On sequential MNIST, the Selective SSM's state-based
training proved memory-intensive, limiting the model sizes we could
compare on our hardware; experiments with larger Selective SSM models
could better isolate architectural contributions.





\section*{Impact statement}
This paper builds upon and draws inspiration from the remarkable works
that introduced State Space Models to the machine learning community,
a class of architectures that have since attracted significant
interest. Our objective is to explore these architectures through the
lens of control theory, a field where State Space Models and dynamical
systems have long been fundamental. By bridging insights from control
systems and machine learning, we aim to contribute to the development
and understanding of new architectures that integrate principles from
both disciplines. While our work has the potential to impact a variety
of applications, we do not find any immediate societal consequences
that we feel must be specifically highlighted at this stage.

\bibliography{New,Main,alias,FP}

@inproceedings{KMC-VL-DD-XS-AG-TS-PH-JQD-AM-LK-DBB-LJC-AW:21,
title={Rethinking Attention with Performers},
author={K. M. Choromanski and V. Likhosherstov and D. Dohan and X. Song and A. Gane and T. Sarlos and P. Hawkins and J. Q. Davis and A. Mohiuddin and L. Kaiser and D. B. Belanger and L. J. Colwell and A. Weller},
booktitle={International Conference on Learning Representations},
year={2021}
}

@InProceedings{AK-AV-NP-FF:21,
  title = 	 {Transformers are {RNN}s: Fast Autoregressive Transformers with Linear Attention},
  author =       {A. Katharopoulos and A. Vyas and N. Pappas and F. Fleuret},
  booktitle = 	 {Proceedings of the 37th International Conference on Machine Learning},
  pages = 	 {5156--5165},
  year = 	 {2020},
}

@inproceedings{RH-ML-TW-MC-AA-DR:23,
title={Liquid Structural State-Space Models},
author={R. Hasani and M. Lechner and T.-H. Wang and M. Chahine and A. Amini and D. Rus},
booktitle={The Eleventh International Conference on Learning Representations },
year={2023}
}

@inproceedings{TD-AG:24,
title={Transformers are {SSM}s: Generalized Models and Efficient Algorithms Through Structured State Space Duality},
author={Tri Dao and Albert Gu},
booktitle={Forty-first International Conference on Machine Learning},
year={2024},
}

@inproceedings{LZ-AS-YD-ASG-YS-BB-MT-AA-SS:24,
  title={{B'MOJO}: Hybrid State Space Realizations of Foundation Models with Eidetic and Fading Memory},
  author={L. Zancato and A. Seshadri and Y. Dukler and A. S. Golatkar and Y. Shen and B. Bowman and M. Trager and A. Achille and S. Soatto},
  booktitle={Advances in Neural Information Processing Systems},
  volume={37},
  pages={130433--130462},
  year={2024}
}

@inproceedings{NMC-AO-BW-CS-TL:24,
  title={Theoretical foundations of deep selective state-space models},
  author={N. Muca Cirone and A. Orvieto and B. Walker and C. Salvi and T. Lyons},
  booktitle={Advances in Neural Information Processing Systems},
  volume={37},
  pages={127226--127272},
  year={2024}
}

@article{GEB:90,
  title={Prefix sums and their applications},
  author={G. E. Blelloch},
  year={1990},
  publisher={School of Computer Science, Carnegie Mellon University Pittsburgh, PA, USA}
}

@Book{GB-GM:91,
  author =	 {G. Basile and G. Marro},
  title =	 {Controlled and Conditioned Invariants in Linear System
                  Theory},
  publisher =	 PH,
  year =	 1991,
  ISBN =	 0131729748,
}

@inproceedings{GM-FM-LN-DM:10,
  title =	 {Geometric Control Theory for Linear Systems: a Tutorial},
  author =	 {Marro, Giovanni and Morbidi, Fabio and Ntogramatzidis,
                  Lorenzo and Prattichizzo, Domenico},
  booktitle =	 mtns,
  novolume =	 5,
  nonumber =	 9,
  year =	 2010,
  pages =	 {1579-1590},
  month =	 jul,
  address =	 {Budapest, Hungary},
}

@book{WMW:85,
  author =	 {W. M. Wonham},
  title =	 {Linear Multivariable Control: A Geometric Approach},
  year =	 1985,
  publisher =	 sv,
  edition =	 3,
  isbn =	 0387960716,
}

@book{CLP-HTN:98,
title = {Digital control system analysis and design (3rd ed.) },
author = {C. L. Phillips and H. T. Nagle},
publisher = {Prentice-Hall, Inc.},
year = {1995},
ISBN={013309832X}
}

@inproceedings{NCR-CE:21,
  title={Parallelizing Legendre Memory Unit Training},
  author={N. C. Reddy and C. Eliasmith},
  booktitle = {Proceedings of the 38th International Conference on Machine Learning},
  year={2021},
  month = {Jul.}
}

@article{SRD-SKS-BBC:22,
title = {Activation functions in deep learning: A comprehensive survey and benchmark},
author = {S. R. Dubey and S. K. Singh and B. B. Chaudhuri},
journal = {Neurocomputing},
year = {2022},
volume = {503},
pages = {92-108}
}

@inproceedings{TM-KC-GSC-JD:13,
  title={Efficient Estimation of Word Representations in Vector Space},
  author={T. Mikolov and K. Chen and G. S. Corrado and J. Dean},
  booktitle = {International Conference on Learning Representations },
  year={2013},
  month = {May}
}

@inproceedings{TM-IS-KC-GC-JD:13,
  title={Distributed representations of words and phrases and their compositionality},
  author={T. Mikolov and I. Sutskever and K. Chen and G. Corrado and J. Dean},
  booktitle = {Proceedings of the 27th International Conference on Neural Information Processing Systems },
  year={2013},
  month = {Dec.}
}

@article{YT-MD-DB-DM:22,
  title={Efficient Transformers: A Survey},
  author={Y. Tay and M. Dehghani and D. Bahri and D. Metzler},
  journal = {ACM Computing Surveys},
  year={2022},
  volume={55},
  number={6},
  pages={28}
}

@inproceedings{TD-DYF-KKS-AWT-AR-CR:23,
  title={Hungry Hungry Hippos: Towards Language Modeling with State Space Models},
  author={T. Dao and D. Y. Fu and K. K. Saab and A. W. Thomas and A. Rudra and C. R{\'e}},
  booktitle = {Proceedings of the 11th International Conference on Learning Representations},
  year={2023},
  month = {May}
}

@misc{JC-ES:24,
  title={Induction Heads as an Essential Mechanism for Pattern Matching in In-context Learning},
  author={J. Crosbie and E. Shutova},
  archivePrefix={arXiv},
  year={2024}
}

@inproceedings{DB-KC-YB:14,
  title={Neural Machine Translation by Jointly Learning to Align and Translate},
  author={D. Bahdanau and K. Cho and Y. Bengio},
  booktitle = {Proceedings of the International Conference on Machine Learning},
  year={2014},
  month = {Apr.}
}

@misc{CO-NE-NN-NJ-ND-TH-BM-AA-YB-AC-TC-DD-DG-ZHD-DH-SJ-AJ-JK-LL-KN-DA-TB-JC-JK-SM-CO:22,
  title={In-context Learning and Induction Heads},
  author={C. Olsson and N. Elhage and N. Nanda and N. Joseph and N. DasSarma and T. Henighan and B. Mann and A. Askell and Y. Bai and A. Chen and T. Conerly and D. Drain and D. Ganguli and Z. Hatfield-Dodds and D. Hernandez and S. Johnston and A. Jones and J. Kernion and L. Lovitt and K. Ndousse and D. Amodei and T. Brown and J. Clark and J. Kaplan and S. McCandlish and C. Olah},
  archivePrefix={arXiv},
  year={2022}
}

@inproceedings{RNP-SM-AM-JTHS-RH-ML-QA-CR-HA-SE-TS-MP-AY:24,
  title={State-free inference of state-space models: the transfer function approach},
  author={R. N. Parnichkun and S. Massaroli and A. Moro and J. T.H. Smith and R. Hasani and M. Lechner and Q. An and C. R\'{e} and H. Asama and S. Ermon and T. Suzuki and M. Poli and A. Yamashita},
  booktitle = {Proceedings of the 41st International Conference on Machine Learning},
  year={2024},
  month = {Jul.}
}

@inproceedings{JS-AW-SWL:23,
  title={Simplified State Space Layers for Sequence Modeling},
  author={J. Smith and A. Warrington and S. W. Linderman},
  booktitle = {Proceedings of the 11th International Conference on Learning Representations},
  year={2023},
  month = {May}
}

@article{AG-TD:23,
  title={Mamba: {L}inear-time sequence modeling with selective state spaces},
  author={A. Gu and T. Dao},
  journal={arXiv preprint arXiv:2312.00752},
  year={2023}
}

@inproceedings{DYF-TD-KKS-AWT-AR-CR:22,
  title={Hungry {H}ungry {H}ippos: Towards Language Modeling with State Space Models},
  author={D. Y. Fu and T. Dao and K. K. Saab and A. W. Thomas and A. Rudra and C. R{\'e}},
  booktitle={The 11th International Conference on Learning Representations},
  year={2022}
}

@inproceedings{AG-KG-CR:21,
  title={Efficiently Modeling Long Sequences with Structured State Spaces},
  author={A. Gu and K. Goel and C. R{\'e}},
  booktitle={International Conference on Learning Representations},
  year={2021}
}

@article{AV-NS-NP-JU-LJ-ANG-LK-IP:17,
  title={Attention is all you need},
  author={A. Vaswani and N. Shazeer and N. Parmar and J. Uszkoreit and L. Jones and A. N. Gomez and {\L}. Kaiser and I. Polosukhin},
  journal={Advances in neural information processing systems},
  volume={30},
  year={2017}
}

@book{JPH:18,
  title={Linear systems theory},
  author={Hespanha, J.~P.},
  year={2018},
  publisher={Princeton university press}
}

@string{science = "Science"}

@STRING{mtns = "Mathematical Theory of Networks and Systems"}

@STRING{sv = "Springer"}

@STRING{PH = "Prentice Hall"}

@STRING{princeton = "Princeton University Press"}
\bibliographystyle{icml2025}

\end{document}